\documentclass[pss,fleqn,embeddedheads]{w-art}
\usepackage{times}
\usepackage{w-thm}

\usepackage[]{graphicx}
\begin{document}
\DOIsuffix{theDOIsuffix} 
\Volume{XX}
\Issue{1}
\Copyrightissue{01}
\Month{01}
\Year{2004}

\pagespan{1}{} 

\Receiveddate{\sf zzz} \Reviseddate{\sf October 30, 2004}
\Accepteddate{\sf zzz} \Dateposted{\sf zzz} 

\title[Spin-polarized transport through carbon nanotubes]{Spin-polarized transport through carbon nanotubes}

\author[S. Krompiewski]{S. Krompiewski\footnote{Corresponding
     author: e-mail:
     {\sf stefan@ifmpan.poznan.pl}, Phone: +48\,61\,86\,95\,126, Fax:
     +48\,61\,86\,84\,524}} 
\address 
{Institute of Molecular Physics, Polish Academy
    of Sciences,  ul. M. Smoluchowskiego 17, 60-179 Pozna\'n,
    Poland}

\begin{abstract}
Carbon nanotubes (CNT) belong to the most promising new materials
which can in the near future revolutionize the conventional
electronics. When sandwiched between ferromagnetic electrodes, the
CNT behaves like a spacer in conventional spin-valves, leading
quite often to a considerable giant magneto-resistance effect
(GMR). This paper is devoted to reviewing some topics related to
electron correlations in CNT. The main attention however is
directed to the following effects essential for electron transport
through nanotubes: (i) nanotube/electrode coupling and (ii)
inter-tube interactions.
 It is shown that these effects may account for
 some recent experimental reports on GMR, including those on
 negative (inverse) GMR.
\end{abstract}
\maketitle                   

\section{Introduction}

A particularly challenging class of materials for the future
electronics are carbon nanotubes, i.e. graphite sheets rolled up
in such a manner that they form very long and thin (in atomic
scale) cylinders. Their length can be of the order of several
micrometers whereas typical cross-sections of single wall carbon
nanotubes (SWCNT) are a few nanometers in diameter, and roughly up
to 10 times more for multi-wall carbon nanotubes (MWCNT) . Both
the SWCNTs and MWCNTs have been attracting much attention of many
researchers since they were discovered \cite{ijima}.

It is now clear that the electronics based on conventional
semiconducting devices (e.g. silicon technology) has faced
fundamental limitations as far as further miniaturization is
concerned. The reasons are various, including the fact that the
relative amount of surface states - destructive for electronic
transport properties - increases rapidly in 2- and 3-dimensional
systems when their sizes are being reduced. Likewise, it is
predicted that the SiO$_2$ layer in the MOSFET will be too thin
soon (within 10 years or so) to retain its insulating properties
\cite{schulz}. So, there is much interest of physicists,
technologists and engineers in new generation nano-scale devices,
including molecular ones. The adopted approach is the so-called
bottom up philosophy, aiming at designing electronic devices at
the molecular level. In contrast to silicon-based devices, the
molecular ones (incl. carbon nanotubes, CNT) can successfully cope
with the more and more demanding miniaturization requirements
without worsening their multi-functional properties. Most probably
the CNTs are also good candidates for spintronics applications in
view of their unusually long spin diffusion length. It is
well-known that electrons can travel through carbon nanotubes up
to many hundreds of nanometers, still keeping their momentum and
spin orientation. So CNTs are believed to be useful in the near
future microelectronics (or molecular electronics) as
interconnects, and possibly also as active elements in integrated
circuits. Physical properties of the CNTs connected to metallic
electrodes depend critically on a quality of CNT/metal junctions,
and may evolve with increasing transparencies of the junctions:
from the Coulomb blockade regime, through the Kondo effect, and
Fabry-Perot resonator-like behavior up to the Fano resonance.

This paper is organized as follows. First, some rather well-known
problems concerning electron correlation regimes in CNTs are
recalled (sub-sections 2.1 - 2.4). Second, a short survey of
selected papers, devoted to the problem of both the CNT-electrode
coupling (Sec. 3) as well as the GMR effect (Sec. 4) will be made.
Finally a tight-binding approach combined with the Green's
function method will be described and applied to the coherent
spin-polarized electronic transport through the CNTs.

\section{Transport regimes}

\subsection{\bf Ballistic regime}

Well-contacted long SWCNTs are essentially one dimensional systems
from the point of view of their transport properties, since they
can be viewed as translational in the axial direction and
quantized in the transverse one. It was shown that in such a case
they behave like a Fabry-Perot resonator, revealing spectacular
$dI/dV-V-V_g $ (differential-conductance vs. bias and gate
voltage) contour plots \cite{liang, PRB02}. The relevant energy
scale here is the energy level spacing $\Delta=h v_F/(2L)$, which
determines the electron wavelength and is directly related to
multiple reflections and superpositions corresponding to the round
trip length 2L of electronic waves, where L and $v_F$ stand for
the SWCNT length, and the Fermi velocity, respectively.

\subsection{\bf Coulomb Blockade}
The opposite transport regime is the Coulomb blockade one. It
takes place at low temperatures when the CNT is weakly coupled to
the electrode and may be regarded as a quantum dot. The relevant
energy scale in this case is the so-called addition energy
($E_{add}$) equal to the inter-level energy spacing ($\Delta E=
\Delta /2$ due to lifting of the band degeneracy) plus the
charging energy $E_c=e^2/C$, where $C=C_s+C_d+C_g$ is the total
capacitance between the dot and the source, drain and gate. In
order for the effect to be pronounced, $E_c$ must be greater than
$\Delta E$, and the junction resistances should be greater than
the inverse of the conductance quantum ($R > h/e^2$). The set of
equations describing the Coulomb blockade reads
\cite{kouwenhoven}:

\begin{equation}
  \label{eq:1}
U(N) = [-e(N - N_0) - C_g V_g]^2/2C,\; \; \; \; F(N) = U(N) +
\sum_{i=1}^{N}E_i,
\end{equation}
\begin{equation}
  \label{eq:3}
\mu_{dot}(N)=F(N) - F(N-1) ,\; \; \; \;E_{add} = \mu_{dot}(N +
1)-\mu_{dot}(N)= E_c +\Delta E,
\end{equation}

where $U(N)$ is the electrostatic energy in the presence of
$N-N_0$ excess electrons and the gate voltage $V_g$, F(N) is the
total energy, and $\mu_{dot}$ and $E_i$ are the chemical potential
and the energy levels of the dot, respectively. In general the
addition energy gap prevents electrons from jumping to and off the
dot. The situation changes when, due to the gate voltage, the
unoccupied energy level of the dot aligns with the chemical
potentials of the contacts (falls into the transport window). The
electrons can then be pumped one by one from source to drain. Such
a process is called sequential tunnelling. There is a simple rule
of thumb to evaluate $E_c$ and $\Delta E$ for SWCNT, namely $E_c
\sim$ 5 meV/ L[$\mu$m] and $\Delta E \sim$ 1 meV/ L[$\mu$m]. So
the ratio $E_c/\Delta E$ is roughly equal to 5 and it does not
depend on the length. The simplest way of energy shell filling in
SWCNT is the 2-fold one (even-odd). Adding an electron to the
occupied level costs just $E_c$ energy, whereas the energy
required to add another electron is increased by what is needed to
reach the successive energy level and amounts to $E_c+\Delta E$.
In some SWCNTs the even-odd shell filling was experimentally
established \cite{cobden}, whereas in others the four-fold shell
filling was reported \cite{liangPRL}.

\subsection{\bf Kondo effect}
It turns out that if the dot/electrodes coupling is not too poor,
a higher order tunnelling process can overcome the Coulomb
blockade. It is called the co-tunnelling and may be thought of as
a correlated jump of an electron from the dot to the drain when
another electron jumps on the dot from the source at the same
time. In the Kondo regime the crucial role is played by processes
for which the entering electron and the leaving one have opposite
spins. The net effect then is as if the quantum dot flipped its
spin, very much like in the conventional (magnetic impurity) Kondo
effect with resonance scattering of conduction electrons by
localized magnetic moments. In the former case however resonance
conductance rather than resonance scattering takes place, and
consequently the conductance increases when temperature is lowered
from the Kondo temperature to zero in contrast to the
aforementioned conventional case. Experimentally, the Kondo effect
is nicely seen in the $dI/dV-V-V_g$ (diamond) plots, as clear
nonzero conductance horizontal features, in the vicinity of V = 0,
within the diamonds corresponding to odd numbers of electrons
\cite{nygard}.

\subsection{\bf Luttinger liquid}
One-dimensional systems with strong electron correlations do not
behave like the Fermi-liquid with free electron-like low-energy
excitations (quasi-particles). Instead, electrons form a
correlated state called the Luttinger liquid (LL) with
plasmon-like excitations. Metallic carbon nanotubes belong to this
class of materials under certain circumstances, and they represent
the best realization of the LL physics \cite{bockrath} . The
strength of electron-electron interactions is described by the
Luttinger parameter $g=1/\sqrt{(1+2 E_c/\Delta E)}$, which is less
than 1 for repulsive Coulomb interactions and equals 1 in the
absence of interactions. The parameter $g$ is just the ratio of
$v_F$ over $v_g$, with the latter being the plasmon velocity.
Incidently, the $v_g$-s for spin plasmons and charge plasmons are
different, but no spin charge separation has ever been
experimentally established in  SWCNTs, to our knowledge. The
interactions critically influence electronic transport, because
the tunnelling of electrons from a metal electrode (Fermi liquid)
to the SWCNT (LL) is only possible if the tunnelling electrons
have energies high enough to couple with the plasmon modes in the
LL. This is in contrast to the tunnelling between two Fermi
liquids, where no energy dependence is basically expected.
Consequently, LL systems show power-law temperature- and
bias-dependence of the conductance
 $ {\cal G} \sim T^\eta$, \;  $dI/dV \sim V^\eta$ ,
for small and large $V$, respectively. Moreover it can be shown
that the scaled differential conductance $\frac{dI}{dV}/T^\eta$ is
a universal function of $eV/k_BT$.

\section{\bf Some experimental aspects}

It is a usual practice to select carbon nanotubes for further
transport measurements on the basis of their room temperature
two-terminal conductance ${\cal G_{RT}}$. Tubes with conductances
almost independent of gate voltage are classified as metallic,
others - as semiconducting. It is convenient to define a
transmission probability of contacts as $P_c=2/(1+{\cal
G}_{max}/{\cal G}_{RT})$, where ${\cal G}_{max}=4 e^2/h$ is the
maximum possible conductance of a SWCNT ($0 \le P_c \le 1 $)
\cite{nygard1}. At low temperature, for low-transparency samples
($P_c \sim 0.15$), Coulomb blockade sets in, with periodic sharp
peaks separated by zero-conductance valleys in the $\cal G$ vs.
$V_g$ plots. For better transparencies $(P_c \sim 0.6)$ the Kondo
regime is reached. Eventually, for $P_c \sim 0.9$ one enters a
ballistic regime, characterized by the so-called inverse diamond
structures with maxima of the conductance near the middle of the
diamonds. It means that very well-contacted SWCNT act as
etalon-like resonant cavities with the (inverted) diamond structue
period varying as $1/L$, which lends support to the view that
open-contact SWCNT are ballistic phase coherent wires. Quite
recently a series of experimental papers has appeared which shows
that nanotubes well-contacted to metal electrodes can now be
fabricated in a controllable way (\cite{javey}, \cite{babic1}).
The best transparent interfaces are for Pd and Au electrodes. In
some cases the transparency is considerably improved upon
annealing (e.g. for Ti electrodes). As shown in \cite{babic2} it
is even possible to observe practically all the transport regimes
together on the same sample by just changing gate voltage. It
means that gate voltage may strongly influence the dot/electrode
tunnelling coupling.

\section{GMR in carbon nanotubes}

The giant magnetoresistance (GMR) effect in CNTs was first
measured 5 years ago \cite{tsukagoshi}, it was estimated that the
spin diffusion length of electrons flowing through a tube is ca.
130 nm or most probably more than that. Bearing in mind, that
nowadays CNT lengths used in electric current measurements, are
quite often as short as 200-250 nm, it means that practically
electrons travel through nanotubes in a spin-coherent way (no spin
flips). Experimental papers concern mostly MWCNTs electrically
contacted by cobalt
 \cite{tsukagoshi,zhao}, but there are also reports on iron
 \cite{jensen} and permalloy \cite{kim} contacted multi-walled and
single wall carbon tubes. The results depend very strongly on
ferromagnetic-electrode/CNT interfaces, yielding the net GMR
effect of the order of 10 - 40\%. It should be stressed however
that also inverse GMR has been reported of roughly similar
magnitude but with negative sign \cite{zhao, kim}. A very
surprising data concerning Fe-contacted SWCNT were reported in
\cite{jensen}, where a measured GMR effect approached 100\%, i.e.
transport was completely blocked in the antiparallel aligned
configuration. It seems that so far a consensus has been gained
that the main contribution to the spin-dependent transport comes
from a ferromagnet/CNT junction and a CNT itself acts only as a
quasi ballistic waveguide. The junction properties are typically
temperature, magnetic field and gate voltage dependent. There are
suggestions that at the interface an antiferromagnetic (e.g. CoO)
layer may be formed, such a layer would be undoubtedly temperature
and magnetic field sensitive, and could also be responsible for
hysteretic (magnetic field sweeping direction) effects. In
\cite{JPCM04} it was shown theoretically that an additional
magnetic layer at the interface does really influence the GMR
effect very substantially, and can even block the current in the
antiparallel configuration. It is also quite possible that local
magnetic domains which touch the carbon nanotube might be
misaligned with respect to the total electrode magnetization.
Other important points concern the internal structure of MWCNT,
i.e first of all (i) the inter-tube interactions, and (ii) whether
or not the particular tubes are in (out of) contact with the
electrodes. These problems will be addressed in the following.

\section{Theoretical approach}

The studies are carried out within the framework of the
single-band tight-binding model for $p_z$ ($\pi$)  electrons in
the carbon nanotubes. The magnetic electrodes are spin-split by
assuming spin-dependent on-site potentials therein. The Green's
function technique has been used, along the line of the
non-equilibrium transport formalism, and under constraint of
global charge neutrality. The so-called extended molecule concept
is adapted, by incorporating to the central part of the system not
only the entire molecule (CNT) but also two closest magnetic
atomic planes from the left- and right-hand side electrodes
\cite{PRB04}. The Green's function is defined as $G =( \hat 1 E -
H_{\rm C}-\Sigma_{\rm L}-\Sigma_{\rm R})^{-1}$, whereas the
density matrix of electrons and current (per spin) are given by:
\begin{equation} \label{rho}
n= \frac{1}{2 \pi} \int dE \, G \left[
 f_L \Gamma_L+f_R \Gamma_R \right] G^\dagger,
 \end{equation}

\begin{equation} \label{I}
I= \frac{e}{h} \int_{- \infty}^\infty dE (f_L-f_R) Tr[\Gamma_L G
\Gamma_R G^\dagger].
\end{equation}

 At low bias and temperature the latter leads to the conductance
${\cal G}= \frac{e^2}{h} Tr\, \left[ \Gamma_L \;G\;\Gamma_{R}
\;G^\dagger \right],$
 where $\alpha=$L, R, C refer to the left, right electrodes and the extended
 molecule, respectively. Moreover
 $\Gamma_\alpha = i ( \Sigma_\alpha - \Sigma_\alpha^\dagger ), \,
 \, \, \, \Sigma_\alpha=V_{{\rm C},\alpha} \, g_\alpha \, V_{{\rm
 C},\alpha}^\dagger \,$, where $V_{L(R),C}$ describes
 the coupling between the electrodes and the extended molecule,
  and $f_\alpha$ is the Fermi-Dirac
 distribution function. The $g_\alpha$ is the $\alpha$-th electrode
 surface Green's function. The latter has been calculated as in
 \cite{todrov}, but summed over the 2-dimensional Brillouin zone
 (while Fourier transforming back to the real space) by the
 special-k-points method \cite{cun}. The "pessimistic" definition
 is used $GMR= (\cal G_{\uparrow,\uparrow}-\cal
G_{\uparrow,\downarrow})/\cal G_{\uparrow,\uparrow}$, where
 the arrows indicate the relative magnetization alignments of the left
 and right external electrodes.
The inter-tube hopping parameters are taken after \cite{roche} as
\begin{equation} \label{t-inter}
t_{int} = -(t/8) \sum_{\,l,\,j}{\cos\theta_{lj}}
e^{\frac{d_{\,l\,j}-b}{\delta}},
\end{equation}
where $\theta$ is the angle between the $\pi$ orbitals, d is a
relative distance, $t$ stands for the nearest neighbor intra-tube
C-C hopping integral (chosen as energy unit), $\delta =$ 0.45
$\AA$ and b = 3.34 $\AA $.

\section{\bf Modelling of the devices and the results}

The contacts are modelled as bulk fcc-(111) slabs infinite in all
the 3 directions and connected to the CNT through a neck composed
of two finite atomic planes. In my previous papers, very special
geometries of the SWCNT (6,6) armchair and the double-walled CNT
(2,2)@(6,6) were studied 
(see \cite{JPCM04} - \cite{PRB04}
 for details). In those cases it was
possible to construct the end-contacted devices in such a way that
interface carbon atoms were placed exactly above the centers of
equilateral triangles formed by the adjacent transition metal
atoms. Now this construction has been generalized so that in
principle any CNT may be end-contacted, by performing a relaxation
under a given potential, e.g. Lennard-Jones' (see Fig.~1). The
procedure goes as follows: (i) nanotubes are generated using the
codes A-2 and A-3 of \cite{saitoMG}, (ii) the tubes are placed
vertically on the fcc-(111) transition metal contacts, (iii) the
outer tube is kept fixed while the electrodes and the inner tube
are free to rotate and move along the axial direction. The
relaxation is made by
 the exact numerical minimization of energy of the device, assuming that
energetically most favorable positions for interface C atoms would
be again those, where they had exactly 3 transition metal atom
nearest neighbors.

 \begin{figure}[htb]
\includegraphics[width=.43\textwidth]{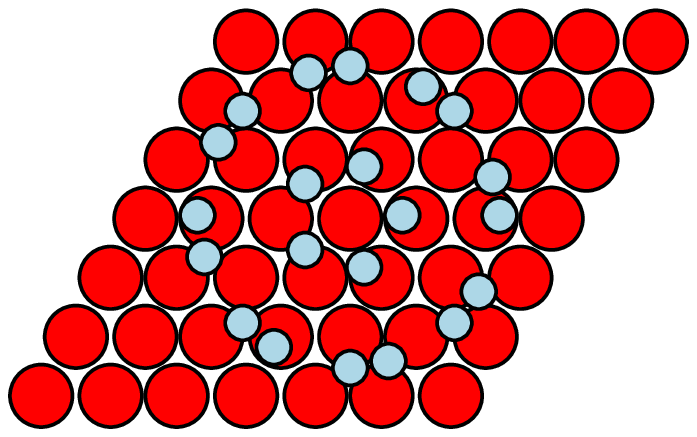}
\hfil
\includegraphics[width=.30\textwidth]{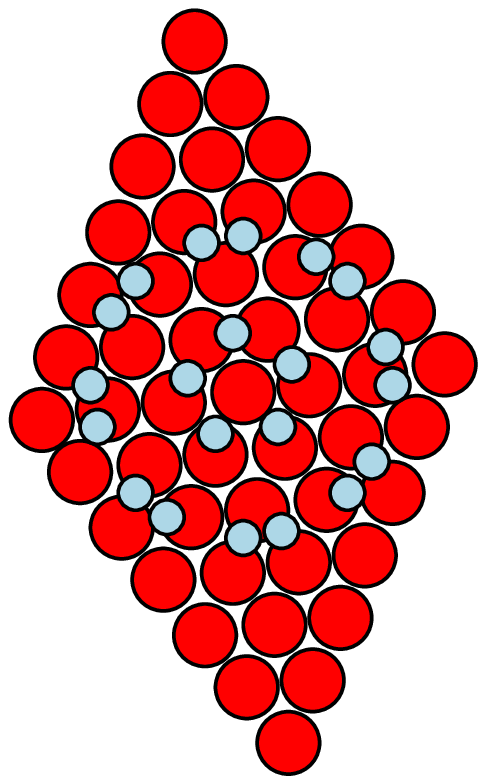}
\caption{One of the interfaces between the DWCNT (5,0)@(8,8)
(small spheres) and the fcc(111) metal electrode (big spheres)
before (left hand side) and after (r.h.s) the relaxation.}
\label{fig:5}
\end{figure}

 In Fig.~2 conductances are presented for parallel (PA - thick solid line) and
 antiparallel (AP - thick dotted line) alignments of magnetic electrodes, along with
 the corresponding GMRs (thin line).
\begin{vchfigure}[htb]
  \includegraphics[width=.75\textwidth]{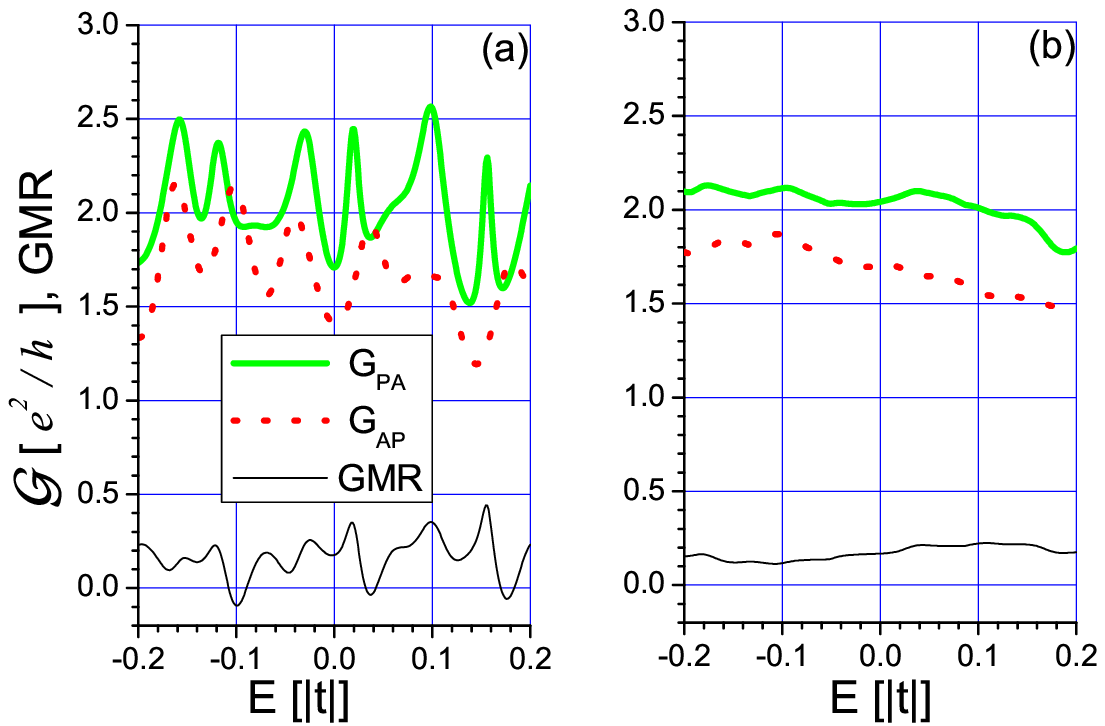}
\vchcaption{(a): Conductance and GMR for a double-walled carbon
nanotube, which consists of 45 zigzag rings and 39 armchair rings
(45-(5,0)@39-(8,8)). (b): The energy averaged conductances over
energy bins equal to the energy-level spacing, and the resulting
GMR. The averaging mimics the anticipated effect of disorder}
\label{fig:13}
\end{vchfigure}
 The figure shows results for a double-walled
 carbon nanotube (DWCNT) composed of the outer armchair (8,8) tube
 and the inner zigzag (5,0) one. In order to keep the length of both the tubes
 close to each other there are 45 and 39 carbon rings for the inner and outer
 tubes, respectively. As readily seen the GMR in the charge neutrality
 point (E = 0) is above 20 \% and happens to be rather robust
 against small energy changes.
 Judging from the
 conductances the presented results correspond to the low
 resistivity junctions. The relevant energy scale is related to
 inter-level spacing $\Delta E \sim 0.15$ in the adopted units ($\left| t \right|
 $ and the graphene lattice constant).
 There are theoretical and experimental
 arguments that such a fine structure gets washed out if there is
 disorder, e.g.  due to interactions with incommensurate
 inner tubes, structural imperfections or impurities \cite{triozon}. Figure~2 (b) shows
 the conductances after averaging over the energy interval
 equal to $\Delta E$, and the resulting GMR.

To gain a deeper insight into the effect of the inner tube on the
overall conductance of the device, Fig.~3  presents the results
for the case when both the tubes are of armchair-type, but the
inner one is by one carbon-ring shorter, being thereby out of
contact to the drain. It is easily seen that now the GMR is quite
unstable with respect to energy changes near $E_F=0$. On the other
hand however, the GMR gets robust and positive if either the
inner-tube interaction vanishes (Fog.~3 (b)) or there are the
above mentioned disorder-induced self-averaging mechanisms. The
present results generalize those of \cite{PRB04}, obtained for
different nanotube diameters and perfect geometical matching at
the interfaces. The GMR of the DWCNT is basically positive, if
\begin{vchfigure}[htb]
\includegraphics[width=
.75\textwidth]{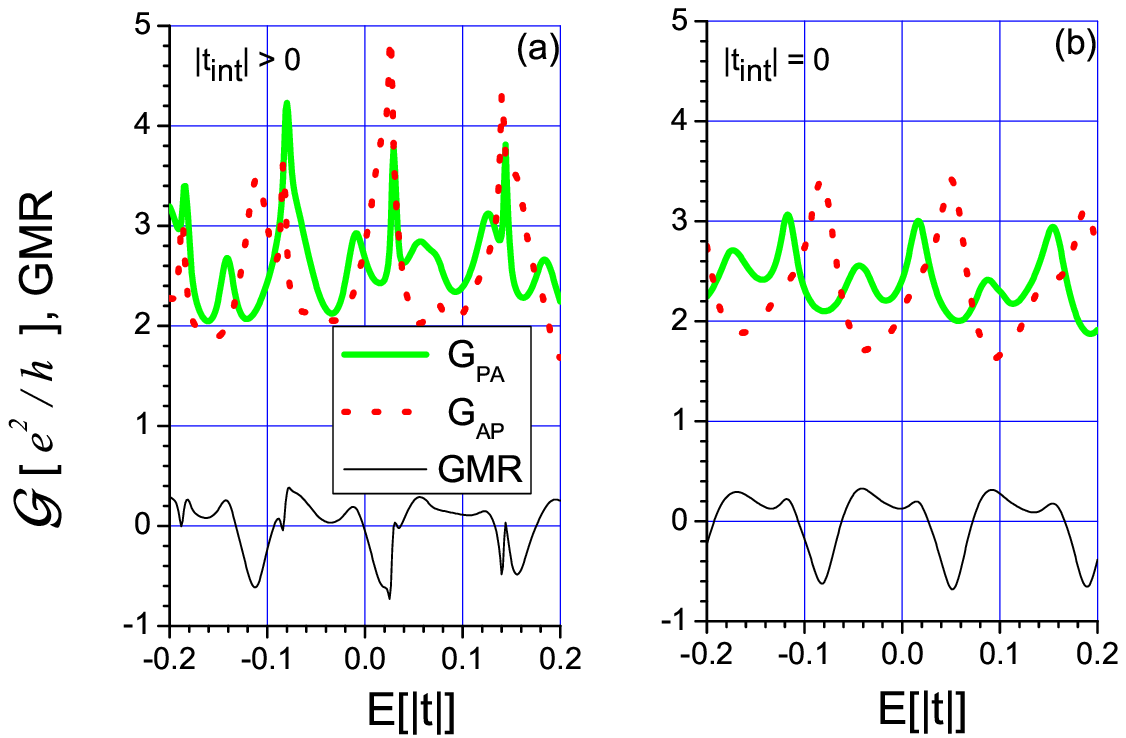} \vchcaption{(a): Conductances and
GMR for the armchair DWCNT (3,3)@(8,8). There are 39 (38) carbon
rings of the outer (inner) tube, so that the inner tube is out of
contact with the drain. (b): Same as in (a) but with no inter-tube
interactions.} \label{fig:12}
\end{vchfigure}
 a wide-spread opinion that the current flows exclusively through the
 outer nanotube is accepted. The new observation is that even in this case
 the inverse GMR may be encountered in the vicinity of $E_F$.
 The inter-tube interactions are quite decisive for electronic
 transport in the DWCNT even if the inner tube is not fully contacted by the
 electrodes, provided it is metallic (of the armchair-type). Note that the
maximum conductance in Fig.~3~(a)
 may exceed $4e^2/h$ in contrast to the cases presented in Figs.~2~(a) and
 3~(b). This means that in the former case the inner tube opens
 additional conduction channels for electric transport.
 Another observation emerging from the present studies is that the inverse GMR
 is more likely to occur in systems with a small amount of disorder,
 where energy-level features do not get smeared out.

\section{\bf Summary}

The spin-dependent transport through carbon nanotubes electrically
contacted by metal electrodes has been addressed. The main
transport regimes, resulting from electron-electron interactions
have been discussed and shown to be strictly related to the
nanotube/electrode junction quality. A method to model CNT/metal
interfaces (end-contacted devices) has been presented and
exemplified by two types of DWCNTs defined in such a way that the
current flows through the outer tube only. It has been shown that
for high-transparency interfaces the GMR effect is basically
positive, the inverse GMR may however be also encountered in this
limit. The latter comes about only on the  inter enegy-level
spacing scale, when due to different boundary conditions, the
conductance has a maximum at the charge neutrality point for the
antiparallel configuration and a reduced value for the parallel
alignment. The probability of such a coincidence is increased by
the fact that on symmetry grounds the conductance peaks are  more
strongly spin-split in the parallel configuration than in the
antiparallel one.


\begin{acknowledgement}
I am grateful to Giovanni Cuniberti for fruitful discussions. The
KBN (research project PBZ-KBN-044/P03-2001) as well as the support
by the Centre of Excellence for Magnetic and Molecular Materials
for Future Electronics, within the European Commission contract
No. G5MA-CT-2002-04049, are also acknowledged.
\end{acknowledgement}


\begin{thebibliography}{10}

\bibitem{ijima} S.~Ijima, Nature (London), {\bf 354}, 56 (1991).
\bibitem{schulz} M.~Schulz, Nature (London), {\bf 399}, 729 (1999).
\bibitem{liang} W. Liang, M. Bockrath, D. Bozovic, J. H. Hafner, M. Tinkham, and H. Park,
Nature (London) {\bf 411}, 665 (2001).
\bibitem{PRB02} S. Krompiewski, J. Martinek, J. Barna{\'s}, Phys. Rev. B, {\bf66}, 073412
(2002).
\bibitem{kouwenhoven} L. P. Kouwenhoven, D. G. Austing, and S.
Tarucha, Rep. Prog. Phys., {\bf 64},~701 (2001).
\bibitem{cobden}  D. H. Cobden and J. Nyg{\aa}rd, Phys Rev. Lett, {\bf 89},~046803 (2002).
\bibitem{liangPRL}  W. Liang, M. Bockrath, and H. Park, Phys Rev. Lett, {\bf 88},~126801 (2002).
\bibitem{nygard} J. Nyg{\aa}rd, D. H. Cobden, and P. E. Lindelof, Nature (London) {\bf 408}, 342 (2000).
\bibitem{bockrath} M. Bockrath, D. H. Cobden, J. Lu, A. G.
Rinzler, R. E. Smalley, L. Balents, and P, L. McEuen, Nature
(London) {\bf 397}, 598 (1999).
\bibitem{nygard1} J. Nyg{\aa}rd, D. H. Cobden, cond-mat/0105289
\bibitem{javey} A. Javey, J. Guo, Q. Wang, M. Lundstrom, and H. Dai,
Nature (London) {\bf 424}, 654 (2003).
\bibitem{babic1} B. Babi\'c, J. Furer, M. Iqbal, and C.
Sch\"onenberger, cond-mat/0406626. 
\bibitem{babic2} B. Babi\'c and C.
Sch\"onenberger, cond-mat/0406571. 
\bibitem{tsukagoshi} K. Tsukagoshi,B.A. Alphenaar, and H. Ago, Nature (London),
{\bf 401}, 572   (1999).
\bibitem{zhao} B. Zhao, I. M{\"o}nch, T. M{\"u}hl, and C.M. Schneider,
Appl. Phys. Lett. {\bf 80}, 3144 (2002).
\bibitem{jensen} A. Jensen, J. Nyg{\aa}rd, and J. Borggreen,
in Toward the controllable quantum states, Proceedings of the
International Symposium on Mesoscopic Superconductivity and
Spintronics, H. Takayanagi and J. Nitta (eds.), pp. 33-37, World
Scientific (2003).
\bibitem{kim} J. Kim, J-R. Kim, J.W. Park, J-J. Kim, K. Kang, N. Kim, B-C. Woo,
Physica E, {\bf 18},~208 (2003).
\bibitem{JPCM04}  S. Krompiewski, J. Phys.: Condens. Matter {\bf 16},~2981 (2004).
\bibitem{krompiewskiJMMM}S. Krompiewski, J. Magn. Magn. Mat. {\bf 1645},~272 (2004).
\bibitem{PRB04}  S. Krompiewski, R. Gutierrez, G. Cuniberti, Phys Rev. B, {\bf
69},~155423 (2004).
\bibitem{todrov} T.N. Todrov et al., J. Phys.: Cond. Mat. {\bf 5}, 2389 (1993)
\bibitem{cun} S.L. Cunningham, Phys. Rev. B {\bf 10}, 4988 (1974)
\bibitem{roche} S. Roche, F. Triozon, A. Rubio, and D. Mayou, Phys. Rev. B {\bf 64},
121401 (2001).
\bibitem{saitoMG}R.~Saito, M.~S.~Dresselhaus, and G.~Dresselhaus,
{\it Physical Properties of Carbon Nanotubes}, Imperial College
Press (1998).
\bibitem{triozon}  F. Triozon, S. Roche, A. Rubio, and D. Mayou, Phys Rev. B, {\bf
69},~121410 (2004).



\end{thebibliography}
\end{document}